\documentclass[12pt]{article}
\usepackage{epsfig}
\normalsize
\def\ba{{\bf a}}
\def\bd{{\bf d}}
\def\bj{{\bf j}}

\def\bp{{\bf p}}
\def\bq{{\bf q}}
\def\bk{{\bf k}}

\def\bz{{\bf z}}
\def\b0{{\bf 0}}
\def\cS{{\cal S}}

\def\bra{\langle}
\def\ket{\rangle}

\def\alf{\alpha}
\def\eps{\epsilon}

\def\lam{\lambda}

\def\sg{\sigma}

\def\det{{\rm det}}
\begin{document}
\title{Reduction formula for fermion loops and density correlations
       of the 1D Fermi gas}
\author{Arne Neumayr and Walter Metzner \\
{\em Institut f\"ur Theoretische Physik C, Technische Hochschule Aachen} \\
{\em Templergraben 55, D-52056 Aachen, Germany}}
\date{\small\today}
\maketitle
\renewcommand{\abstractname}{\normalsize{Abstract}}
\begin{abstract}
Fermion $N$-loops with an arbitrary number of density
vertices $N > d+1$ in $d$ spatial dimensions can be expressed
as a linear combination of $(d+1)$-loops with coefficients that
are rational functions of external momentum and energy variables.
A theorem on symmetrized products then implies that
divergencies of single loops for low energy and small momenta
cancel each other when loops with permuted external variables are
summed.
We apply these results to the one-dimensional Fermi gas, where an
explicit formula for arbitrary $N$-loops can be derived.
The symmetrized $N$-loop, which describes the dynamical $N$-point 
density correlations of the 1D Fermi gas, does not diverge for
low energies and small momenta. 
We derive the precise scaling behavior of the symmetrized $N$-loop 
in various important infrared limits.
\end{abstract}

\vfill\eject

\noindent
{\large\bf 1. Introduction} \par
\medskip

The properties of fermion loops with density vertices (see Fig.\ 1) 
play a role in the theory of Fermi systems and various other
problems in statistical mechanics.
Symmetrized loops, obtained by summing all permutations of the $N$ 
external energy-momentum variables of a single $N$-loop, describe 
dynamical $N$-point density correlations of a (non-interacting)
Fermi gas.
Single loops have no direct physical meaning (for $N>2$), but
contribute as subdiagrams of Feynman diagrams in the perturbation
expansion of interacting Fermi systems.
Symmetrized loops appear as integral kernels in effective actions
for interacting Fermi systems, where fermionic degrees of freedom
have been eliminated in favor of collective density fluctuations
\cite{Pop}.
The behavior of symmetrized loops for small energy and momentum
variables is particularly important for Fermi systems with 
long-range interactions, whose Fourier transform is singular for
small energy and momentum transfers \cite{MCD,Kop}.
\par\smallskip
Besides their relevance for interacting electron systems and other
fermionic systems in nature, the theory of Fermi systems has also
a bearing on various problems in classical statistical mechanics,
which can be mapped to an effective Fermi system (gas or interacting).
For example, the statistical mechanics of directed lines in two
dimensions can be mapped to the quantum mechanics of fermions in
one spatial dimension \cite{PT}. 
This mapping has been exploited extensively to study fluctuations 
of crystal surfaces \cite{Nij,BK}.
\par\smallskip
The 2-loop, corresponding to the 2-point density correlation
function has been computed long ago in one, two, and three
dimensions \cite{FHN}.
Recently, Feldman et al.\ \cite{FKST} have obtained an exact
expression for the $N$-loop with arbitrary energy and momentum
variables in two dimensions. 
We have evaluated that expression explicitly and analyzed the
small energy-momentum limit of the symmetrized loops, showing
in particular that infrared divergencies of single loops cancel
completely in the sum over permutations \cite{NM}.
\par\smallskip
Most recently, Wagner \cite{Wag} has published a reduction formula 
for fermion loops in the static case, where all energy variables
are set zero. 
This formula reduces the $N$-loop for a $d$-dimensional Fermi
system to a linear combination of $(d+1)$-loops, with coefficients
that are rational functions of the momenta.
In this work we point out that Wagner's formula and derivation
can be easily extended to the case of finite energy variables
(Sec.\ 3).
In the two-dimensional case, the possibility of such an extension
is evident from the exact expression for $N$-loops \cite{FKST}.
The small energy-momentum behavior of symmetrized $N$-loops
can be analyzed by applying a theorem on symmetrized products 
derived in our work on two-dimensional systems \cite{NM}, which we 
formulate for the general $d$-dimensional case in Sec.\ 4.
We apply the reduction formula to a one-dimensional system, where
the $N$-loop can be expressed in terms the 2-loop, which
is very easy to compute (Sec.\ 5). We finally compute the infrared
scaling behavior of symmetrized $N$-loops in a one-dimensional
Fermi system.
\par

\vskip 1cm

\noindent
{\large\bf 2. Loops} 
\par
\medskip

The amplitude of the $N$-loop with density vertices, represented by the 
Feynman diagram in Fig.\ 1, is given by
\begin{equation}\label{eq1}
 \Pi_N(q_1,\dots,q_N) = I_N(p_1,\dots,p_N) =  
 \int\! \frac{d^dk}{(2\pi)^d} \int\! \frac{dk_0}{2\pi} \, 
 \prod_{j=1}^N  G_0(k\!-\!p_j)
\end{equation}
at temperature zero.
Here $k = (k_0,\bk)$, $q_j = (q_{j0},\bq_j)$, and $p_j = (p_{j0},\bp_j)$
are $(d+1)$-dimensional energy-momentum vectors.
We use natural units, i.e.\ $\hbar = 1$.
The variables $q_j$ and $p_j$ are related by the linear transformation
\begin{equation}\label{eq2}
 q_j = p_{j+1} - p_j \; , \quad j = 1,\dots,N
\end{equation}
where $p_{N+1} \equiv p_1$. 
Energy and momentum conservation at all vertices
yields the restriction $q_1 + \dots + q_N = 0$.
The variables $q_1,\dots,q_N$ fix $p_1,\dots,p_N$ only up to a constant
shift $p_j \mapsto p_j + p$. Setting $p_1 = 0$, one gets
\begin{eqnarray}\label{eq3}
 p_2 & = & q_1 \nonumber \\
 p_3 & = & q_1 + q_2 \nonumber \\
     & \vdots & \nonumber \\
 p_N & = & q_1 + q_2 + \dots + q_{N-1} \; .
\end{eqnarray}
We use the imaginary time representation, with a 
non-interacting propagator
\begin{figure}
\hspace*{3cm}
\epsfig{file=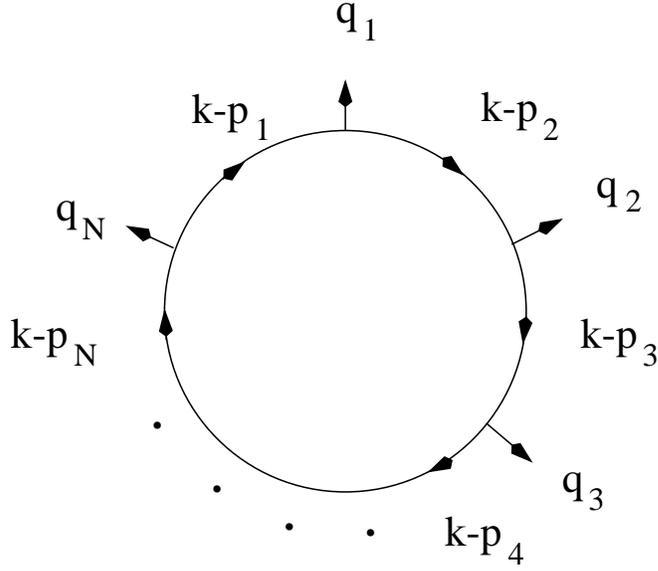, width=8.6cm}
\caption{The N-loop with its energy-momentum labels.}
\end{figure}
\begin{equation}\label{eq4}
 G_0(k) = \frac{1}{ik_0 - (\eps_{\bk} - \mu)} 
\end{equation}
where $\eps_{\bk}$ is the dispersion relation and $\mu$ the chemical
potential of the system.
For a continuum (not lattice) Fermi system the dispersion relation
is $\eps_{\bk} = \bk^2/2m$, where $m$ is the fermion mass.
The $k_0$-integral in Eq.\ (\ref{eq1}) can be easily carried out using the
residue theorem; one obtains \cite{FKST}
\begin{equation}\label{eq5}
 I_N(p_1,\dots,p_n) = \sum_{i=1}^N \int_{|\bk-\bp_i| < k_F}
 \frac{d^dk}{(2\pi)^d} 
 \left( \prod_{j=1 \atop j \neq i}^n f_{ij}(\bk) \right)^{-1}
\end{equation}
where 
$f_{ij}(\bk) = i(p_{i0} - p_{j0}) + \eps_{\bk-\bp_i} - \eps_{\bk-\bp_j}$.
\par\smallskip
The 2-loop $\Pi_2(q,-q) \equiv \Pi(q)$ is known as polarization insertion
or particle-hole bubble, and has a direct physical meaning: $\Pi(q)$ is
the dynamical density-density correlation function of a non-interacting
Fermi system \cite{FW}.
For $N>2$, the $N$-loop is not a physical quantity, but the 
{\em symmetrized}\/ $N$-loop
\begin{equation}\label{eq6}
 \Pi^S_N(q_1,\dots,q_N) = \cS \, \Pi_N(q_1,\dots,q_N) =
 \frac{1}{N!} \sum_P \Pi_N(q_{P1},\dots,q_{PN}) \: ,
\end{equation}
where the symmetrization operator $\cS$ imposes summation over all 
permutations of $q_1,\dots,$ $q_N$, is proportional to the (connected)
dynamical $N$-point density correlation function:
\begin{equation}\label{eq7}
 \bra \rho(q_1),\dots,\rho(q_N) \ket_{con} =
 (-1)^{N-1} (N-1)! \, \Pi^S_N(q_1,\dots,q_N)
\end{equation}
Here $\rho(q)$ is the Fourier transform of the particle density 
operator. Eq.\ (7) is easily verified by applying Wick's theorem 
\cite{FW}.
Note that Wick's theorem yields a sum of $(N \!-\! 1)!$ distinct loops 
with non-equivalent permutations of $q_1,\dots,q_N$, while the sum
in Eq.\ (6) includes cyclic permutations which produce $N$ 
equivalent copies of each loop.
\par

\vskip 1cm

\noindent
{\large\bf 3. Reduction formula} \par
\medskip
We now state the {\em reduction formula}\/ that reduces the $N$-loop
for a $d$-dimensional system with $N > d+1$ to a linear combination
of $(d+1)$-loops with coefficients that are explicitly computable
rational functions of momentum and energy variables.
This formula is a straightforward generalization of a result derived
recently by Wagner \cite{Wag} for the static case $p_{j0} = 0$.
\par\smallskip
Let $p_1,\dots,p_N$ be such that for each tupel of integers 
$\bj = (j_1,\dots,j_{d+1})$ with $1 \leq j_1 < \dots < j_{d+1} \leq N$,
the complex d-dimensional vectors $\bd^{\bj}$ determined by the
linear equations
\begin{equation}\label{eq8}
 f_{j_1j_r}(\bd^{\bj}) = 
 i (p_{j_10} - p_{j_r0}) + 
 \frac{1}{2m} \, (\bp_{j_1}^2 - \bp_{j_r}^2) +
 \frac{1}{m} \, (\bp_{j_r} - \bp_{j_1}) \cdot \bd^{\bj} = 0
\end{equation}
for $r = 2,\dots,d+1$ are well-defined and unique.
Suppose that for $n = 1,\dots,N$ with $n \neq j_1,\dots,j_{d+1}$
the numbers
\begin{equation}\label{eq9}
 f_n^{\bj} := f_{j_r n}(\bd^{\bj}) = 
 i (p_{j_r0} - p_{n0}) + 
 \frac{1}{2m} \, (\bp_{j_r}^2 - \bp_n^2) +
 \frac{1}{m} \, (\bp_n - \bp_{j_r}) \cdot \bd^{\bj}
\end{equation}
are non-zero. Then
\begin{equation}\label{eq10}
 I_N(p_1,\dots,p_N) = 
 \sum_{j_1,\dots,j_{d+1} \atop 1 \leq j_1 < \dots < j_{d+1} \leq N}
 \left( \prod_{n=1 \atop n \neq j_1,\dots,j_{d+1}}^N
 \frac{1}{f_n^{\bj}} \right) \,
 I_{d+1}(p_{j_1},\dots,p_{j_{d+1}})
\end{equation}
Note that the numbers $f_{j_r n}(\bd^{\bj})$ with $r = 1,\dots,d+1$
are all equal, as a consequence of Eq.\ (8). 
The vector $\bd^{\bj}$ is uniquely defined
if the vectors $\bp_{j_r} - \bp_{j_1}$, where $r = 2,\dots,d+1$, are 
linearly independent. 
The real part of $\bd^{\bj}$ is the center of the uniquely defined 
circumscribing sphere through the points $\bp_{j_1},\dots,\bp_{j_{d+1}}$
in $d$-dimensional euclidean space.
In contrast to $\bd^{\bj}$, the numbers $f_n^{\bj}$ are invariant
under a shift $p_j \mapsto p_j + p$ and can thus be expressed in 
terms of the variables $q_1,\dots,q_N$.
\par\smallskip
The proof of the above reduction formula, a simple generalization of
the proof given by Wagner \cite{Wag} for the static case, is presented
in the Appendix.

\vskip 1cm

\noindent
{\large\bf 4. Symmetrized products} \par
\medskip

Symmetrized loops are obtained by summing over all permutations
of external energy-momentum variables $q_1,\dots,q_N$ as in 
Eq.\ (6).
Up to a trivial constant, symmetrized $N$-loops are the connected 
$N$-point density correlation functions of the Fermi gas.
The behavior of these functions in the infrared limit $q_j \to 0$
determines the long-distance (in space and time) density correlations, 
and is a crucial ingredient for power-counting of contributions to 
effective actions for collective density fluctuations.
We will consider two important scaling limits: \\
i) {\em small energy-momentum}\/ limit 
 $\lim_{\lam \to 0} \Pi^S_N(\lam q_1,\dots,\lam q_N)$ , \\
ii) {\em dynamical}\/ limit
 $\lim_{\lam \to 0} 
  \Pi^S_N[(q_{10},\lam \bq_1),\dots,(q_{N0},\lam \bq_N)]$ . \\
Single $N$-loops diverge generally (for almost all choices of 
$q_1,\dots,q_N$) as $\lam^{2-N}$ in the small energy-momentum limit, 
which is what one would expect from simple power-counting applied 
to the integral (1).
A notable exception is the socalled {\em static}\/ limit, where the 
momenta $\bq_j$ tend to zero after all energy variables $q_{j0}$
have vanished. In that case one obtains a unique finite limit
$\Pi_N \to \frac{(-1)^{N-1}}{(N-1)!} \frac{d^{N-2}}{d\eps^{N-2}}
 \left. D(\eps) \right|_{\eps=\mu}$, where $D(\eps)$ is the density 
of states \cite{HK}.
In the following we will show that systematic cancellations occur in 
the sum over permutations in the general small energy-momentum limit
and also in the dynamical limit.
\par\smallskip
The factor multiplying the $(d+1)$-loops in the reduction formula
can be written as 
\begin{equation}\label{eq11}
 {\cal F}^{\bj} := 
 \prod_{n=1 \atop n \neq j_1,\dots,j_{d+1}}^N 
 \frac{1}{f_n^{\bj}} =
 \prod_{r=1}^{d+1} F_r^{\bj}
\end{equation}
where
\begin{equation}\label{eq12}
 F_r^{\bj} = \left\{ \begin{array}{lcr}
 F_r^{\bj}(q_{j_r},q_{j_r+1},\dots,q_{j_{r+1}-1}) = 
 \prod_{n=j_r+1}^{j_{r+1}-1} \frac{1}{f_n^{\bj}}
   & \mbox{for} & j_{r+1} > j_r + 1 \\
 1 & \mbox{for} & j_{r+1} = j_r + 1
 \end{array} \right.
\end{equation}
Here $j_{d+2} \equiv j_1$, i.e.\ for $r=d+1$ the index $n$ runs 
from $j_{d+1}+1$ to $N$ and then from $1$ to $j_1-1$. 
Note that $F_r^{\bj}$ depends also on differences of the 
energy-momentum variables $p_{j_1},\dots,p_{j_{d+1}}$, besides the 
explicitly written arguments.
As a product of $M_r = j_{r+1} - j_r -1$ factors $(f_n^{\bj})^{-1}$, 
$F_r^{\bj}$ diverges as $\lam^{-M_r}$ in the small energy-momentum
limit, since each $f_n^{\bj}$ vanishes linearly.
We define a symmetrized product
\begin{equation}\label{eq13}
 S_r^{\bj}(k_1,\dots,k_{M_r+1}) =
 \frac{1}{(M_r \!+ 1)!} \sum_P  
 F_r^{\bj}(k_{P1},\dots,k_{P(M_r+1)})
\end{equation}
where all permutations of $k_1,\dots,k_{M_r+1}$ are summed.
According to the following theorem, the symmetrized product 
$S_r^{\bj}$ can be expressed such that the cancellations of 
singularities in the infrared limit become obvious.
\par\smallskip
\noindent
{\em Factorization theorem}:
The symmetrized product $S_r^{\bj}$ can be written as
$\frac{m^{M_r}}{(M_r+1)!}$ times a sum over fractions with 
numerators
\begin{equation}\label{eq14}
 (\bk_{\sg_1} \!\cdot \bk_{\sg'_1}) 
 (\bk_{\sg_2} \!\cdot \bk_{\sg'_2})
 (\bk_{\sg_{M_r}} \!\cdot \bk_{\sg'_{M_r}})
\end{equation}
where $\sg_i \neq \sg'_i$ and $M_r = j_{r+1}-j_r-1$, 
and products of $2M_r$ functions $f^{\bj}$ as denominators.
The functions $f^{\bj}$ have the form
\begin{equation}\label{eq15}
 f^{\bj}(p,p') = i(p_0 - p'_0) + 
 \frac{1}{2m} (\bp^2 - \bp'^2) + (\bp' - \bp) \cdot \bd^{\bj}
\end{equation}
where $p = p_{j_r}$ and 
$p' = p_{j_r} + (\mbox{\em partial sum of} \; k_1,\dots,k_{M_r+1})$.
In each numerator, each momentum variable $k_1,\dots,k_{M_r+1}$ 
appears at least once as a factor in one of the scalar products.
\par\smallskip
\noindent
For example, in the simplest case $M_r = 1$ one obtains
\begin{equation}\label{eq16}
 F_r^{\bj}(k_1,k_2) + F_r^{\bj}(k_2,k_1) =
 \frac{m \, (\bk_1 \!\cdot \bk_2)}
  {f^{\bj}(p_{j_r},p_{j_r}\!+k_1) \, 
   f^{\bj}(p_{j_r},p_{j_r}\!+k_2)}
\end{equation}
The factorization theorem has been derived recently \cite{NM}
in the context of two-dimensional systems. 
The proof provides a concrete algorithm leading to the factorized
expression.
Since the algorithm is actually independent of the dimensionality 
of the system, we will not repeat the derivation here.
\par\smallskip
The infrared scaling behavior of $S_r^{\bj}$ follows directly:
\par\smallskip
\noindent
i) $S_r^{\bj}$ is {\em finite}\/ (of order one) and {\em real}\/
in the small energy-momentum limit.
\par\smallskip
\noindent
ii) $S_r^{\bj}$ vanishes as $\lam^{2M_r}$ in the dynamical limit.
\par\smallskip
\noindent
To see this, note that the functions $f^{\bj}(p,p')$ vanish linearly 
in the small energy-momentum limit, and are purely imaginary to
leading order in $\lam$, while they remain finite in the dynamical 
limit.
\par\smallskip
The symmetrized product is thus much smaller for small energy and
momentum variables than each single term, namely by a factor
$\lam^{M_r}$ in the small energy-momentum limit, and even by a
factor $\lam^{2M_r}$ in the dynamical limit. 
This result holds in any dimension $d$.
\par

\vskip 1cm

\noindent
{\large\bf 5. One-dimensional systems} \par
\medskip 

We now apply the general results from Secs.\ 3 and 4 to one-dimensional
systems \cite{Voi}, where particularly simple expressions can be 
obtained.
We consider first single, then symmetrized loops.
\par
\bigskip

\noindent
{\bf A) Single Loops} \par
\medskip

In one dimension, the reduction formula (10) reduces 
$N$-loops to linear combinations of 2-loops:
\begin{equation}\label{eq17}
 I_N(p_1,\dots,p_N) = 
 \sum_{j_1,j_2 \atop 1 \leq j_1 < j_2 \leq N}
 \left[ \prod_{n=1 \atop n \neq j_1,j_2}^N 
 \frac{1}{f_n^{\bj}}
 \right] \, I_2(p_{j_1},p_{j_2})
\end{equation}
where $d^{\bj}$ is given explicitly by
\begin{equation}\label{eq18}
 d^{\bj} = \frac{1}{2} (p_{j_1 1} + p_{j_2 1}) +
 im \, \frac{p_{j_1 0} - p_{j_2 0}}{p_{j_1 1} - p_{j_2 1}}
\end{equation}
and
\begin{equation}\label{eq19}
 f_n^{\bj} = 
 - \frac{1}{2m} (p_{n1} - p_{j_1 1})(p_{n1} - p_{j_2 1}) +
 i(p_{j_1 0} - p_{n0}) + i(p_{n1} - p_{j_1 1}) \,
 \frac{p_{j_1 0} - p_{j_2 0}}{p_{j_1 1} - p_{j_2 1}}
\end{equation}
Here $p_{n1}$ and $p_{j_r 1}$ are the one-dimensional momentum
components of the energy-momen\-tum vectors $p_n = (p_{n0},p_{n1})$
and $p_{j_r} = (p_{j_r 0},p_{j_r 1})$, respectively.
The 2-loop can be computed very easily, the result being
\begin{equation}\label{eq20}
 I_2(p_{j_1},p_{j_2}) = \frac{m}{\pi} \, 
 \frac{1}{p_{j_1 1} - p_{j_2 1}} \,
 \log \left| \frac{k_F - \alf_{j_1 j_2}}{k_F + \alf_{j_1 j_2}} \right|
\end{equation}
where 
\begin{equation}\label{eq21}
 \alf_{j_1 j_2} = 
 \frac{1}{2} \, (p_{j_1 1} - p_{j_2 1}) + 
 im \, \frac{p_{j_1 0} - p_{j_2 0}}{p_{j_1 1} - p_{j_2 1}}
\end{equation}
We have thus obtained an explicit expression in terms of elementary
functions for $N$-loops in one dimension.
One may easily perform an analytic continuation to real (instead of
imaginary) energy variables, $ip_{j0} \mapsto \eps_j$, in the
above expressions to analyze, for example, the non-linear dynamical
density response of the Fermi gas.
\par\smallskip
In the zero energy limit $p_{j0} \to 0$ one obtains the simple result
\begin{eqnarray}\label{eq22}
 \lim_{p_{j0} \to 0 \atop j = 1,\dots,N} I_N(p_1,\dots,p_N) =
 \hskip 10cm \nonumber \\
 \sum_{j_1,j_2 \atop 1 \leq j_1 < j_2 \leq N}
 \left[ \prod_{n=1 \atop n \neq j_1,j_2}^N  
 \frac{-2m}{(p_{n1} \!-\! p_{j_1 1})(p_{n1} \!-\! p_{j_2 1})}
 \right] \, \frac{m}{\pi(p_{j_1 1} \!-\! p_{j_2 1})} \,
 \log \left|
 \frac{2k_F - (p_{j_1 1} \!-\! p_{j_2 1})}
      {2k_F + (p_{j_1 1} \!-\! p_{j_2 1})}
 \right|
\end{eqnarray}
Note that the above expression has a finite limit for $p_{j1} \to 0$,
although each contribution to the sum diverges.
\bigskip

\noindent
{\bf B) Symmetrized Loops} \par
\medskip

It is well known that for a {\em linearized}\/ dispersion relation
$\eps_k = v_F (|k|-k_F)$, as in the one-dimensional Luttinger
model, the symmetrized $N$-loop $\Pi^S_N(q_1,\dots,q_N)$ vanishes 
identically for $N>2$ even for finite $q_j$ with sufficiently small
momenta $q_{j1}$ \cite{DL}.
We now analyze the infrared behavior of symmetrized $N$-loops in a
one-dimensional system with the usual quadratic dispersion relation.
Symmetrizing the reduction formula, we can write symmetrized loops 
as
\begin{equation}\label{eq23}
 \Pi^S_N(q_1,\dots,q_N) =
 \cS \!\! \sum_{j_1,j_2 \atop 1 \leq j_1 < j_2 \leq N} 
 S_1^{\bj} \, S_2^{\bj} \, I_2(p_{j_1},p_{j_2})
\end{equation}
where $\cS$ is the symmetrization operator introduced in Sec.\ 2
and $S_1^{\bj}$ and $S_2^{\bj}$ are the symmetrized products 
defined in Sec.\ 4.
Note that first symmetrizing partially (with respect to a subset of 
variables, as in the products $S_r^{\bj}$) and then completely
(by applying $\cS$) yields the same result as symmetrizing
everything just once.
\par\smallskip
We can now easily derive the scaling behavior of $\Pi_N^S$ in
the small energy-momentum and dynamical limit, respectively.
The 2-loop $\Pi(q_1) \equiv \Pi_2(q_1,-q_1) = I_2(0,q_1)$ tends to the
finite value
\begin{equation}\label{eq24}
 \Pi(\lam q_1) \to - \frac{1}{\pi v_F} \,
 \frac{1}{1 + [q_{10}/(v_F q_{11})]^2}
\end{equation}
in the small energy-momentum limit and vanishes quadratically as
\begin{equation}\label{eq25}
 \Pi(q_{10},\lam q_{11}) \to - \frac{v_F}{\pi} \,
 \frac{q_{11}^2}{q_{10}^2} \, \lam^2
\end{equation}
in the dynamical limit, where $v_F = k_F/m$ is the Fermi velocity.
The same behavior is found for the 2-loop with a linearized $\eps_k$.
Since $S_1^{\bj}$ and $S_2^{\bj}$ are both finite in the small
energy momentum limit, the symmetrized $N$-loop remains finite,
too:
\begin{equation}\label{eq26}
 \Pi_N^S(\lam q_1,\dots,\lam q_N) = O(1) \quad
 \mbox{for} \quad \lam \to 0 \; .
\end{equation}
Only in the static case $q_{j0} = 0$ each single loop $\Pi_N$ 
has a finite limit for $q_{j1} \to 0$, while in general the 
above result is due to systematic cancellations of infrared
divergencies.
In the dynamical limit the product $S_1^{\bj} S_2^{\bj}$ vanishes
as $\lam^{2M_1+2M_2}$ where $M_1 + M_2 = N - 2$, such that
\begin{equation}\label{eq27}
 \Pi^S_N[(q_{10},\lam q_{11}),\dots,(q_{N0},\lam q_{N1})] =
 O(\lam^{2N-2}) \quad 
 \mbox{for} \quad \lam \to 0 \; .
\end{equation}
The same scaling behavior has been found previously for 
two-dimensional systems \cite{NM}.
\par

\vskip 1cm

\noindent
{\large\bf 6. Conclusion} \par
\medskip

We have derived a formula that reduces the evaluation of fermion
loops with $N$ density vertices in $d$ dimensions to the 
computation of loops with only $d+1$ vertices.
This was obtained by a straightforward extension of a recent
result by Wagner \cite{Wag} for the zero energy limit to
arbitrary energy variables.
Using a theorem about symmetrized products, we have shown that
infrared divergencies of single loops cancel to a large extent
when permutations of external energy-momentum variables are 
summed. The symmetrized $N$-loop, which is proportional to the
$N$-point density correlation function of the Fermi gas, is
thus generally much smaller in the infrared limit than 
unsymmetrized loops.
For one-dimensional systems, we have obtained an explicit
expression for arbitrary $N$-loops in terms of elementary
functions of the energy-momentum variables. 
We have shown that symmetrized loops do not diverge for
low energies and small momenta. 
In the dynamical limit, where momenta scale to zero at fixed
energy variables, the symmetrized $N$-loop vanishes as the
$(2N \!-\! 2)$th power of the scale parameter.
\par\smallskip
We finally outline some applications of our results.
\par\smallskip
{\em Evaluation of Feynman diagrams:}\/ 
Analytical results for loops are of course useful for computing
Feynman diagrams containing fermion loops as subdiagrams.
The number of energy-momentum variables that remain to be 
integrated (analytically or numerically) is thus reduced.
In particular, the mutual cancellation of contributions 
associated with different permutations of energy-momentum
transfers entering a loop can be treated analytically,
avoiding numerical ``minus-sign'' problems.
\par\smallskip
{\em Effective actions:}\/
Effective actions for interacting Fermi systems, where the fermionic
degrees of freedom have been eliminated in favor of collective
density fluctuations, contain symmetrized $N$-loops as kernels
\cite{Pop}. 
A good control of the infrared behavior of these kernels is 
essential for assessing the relevance of non-Gaussian terms in 
the effective action, especially in the presence of long-range
interactions.
In one-dimensional systems one can use our results to compute
the scaling dimensions of corrections to the leading low-energy 
behavior of Luttinger liquids \cite{Hal} by analyzing the 
non-quadratic corrections in the bosonized action. 
\par\smallskip
{\em Surface fluctuations:}\/
Some models of surface fluctuations lead to the statistical
mechanics of directed lines in two dimensions,
which can be mapped to the quantum mechanics of fermions in one
spatial dimension \cite{Nij,BK}.
Most recently, Pr\"ahofer and Spohn \cite{PS} have shown that
the probability distribution of height fluctuations in such models
is Gaussian at long distances on the surface.
For this result it was enough to establish that symmetrized
$N$-loops in the associated Fermi system are less singular than the
naive power-counting estimate.
Our result Eq.\ (26) yields the precise scaling dimension of 
non-Gaussian terms, and implies in particular that high order 
corrections vanish very rapidly at long distances.
\par

\vskip 1cm


\noindent{\bf Acknowledgments:} \\
We are grateful to H. Kn\"orrer, H. Spohn, and E. Trubo\-witz for 
valuable discussions. 

\vskip 2cm


\noindent
{\large\bf Appendix A: Proof of reduction formula } \par
\medskip

Following Wagner's \cite{Wag} derivation for the static case, we 
prove the reduction formula (10) by applying the following 
many-dimensional version of Lagrange's interpolation formula:
\par\smallskip
\noindent
{\em Lemma:}\/ 
Suppose that $1 \leq d+1 < N$ and the $(d+1)$-dimensional complex
vectors $\ba_1,\dots,\ba_N$ are such that 
$\ba_{j_1},\dots,\ba_{j_{d+1}}$
as well as $(\ba_{j_1}-\ba_n),\dots,(\ba_{j_{d+1}}-\ba_n)$ are linearly
independent for pairwise different indices 
$j_1,\dots,j_{d+1},n \in \{1,\dots,N\}$.
For $\bj = (j_1,\dots,j_{d+1})$ 
with $1 \leq j_1 < \dots < j_{d+1} \leq N$ determine the complex
$(d+1)$-dimensional vector $\bz^{\bj}$ by the system of linear
equations $\ba_{j_r} \!\cdot \bz^{\bj} = 1$ for $r = 1,\dots,d+1$.
Then each complex homogeneous polynomial $P(z_0,z)$ of degree
$N-(d+1)$ in the $d+2$ variables $z_0,\bz=(z_1,\dots,z_{d+1})$ can
be written as
\begin{equation}\label{A1}
 P(z_0,\bz) = 
 \sum_{j_1,\dots,j_{d+1} \atop 1 \leq j_1 < \dots < j_{d+1} \leq N} 
 P(1,\bz^{\bj})
 \prod_{n=1 \atop n \neq j_1,\dots,j_{d+1}}^N
 \frac{(z_0 - \ba_n \cdot \bz) \, \det(\ba_{j_1},\dots,\ba_{j_{d+1}})}
      {\det \left( \begin{array}{cccc}
       1 & 1 & \dots & 1 \\
       \ba_n & \ba_{j_1} & \dots & \ba_{j_{d+1}}
       \end{array} \right) }
\end{equation}
where the vectors $\ba_1,\dots,\ba_N$ enter the determinants as 
column vectors.
\par\smallskip
\noindent
For a proof, see Ref.\ \cite{OW}.
\par\smallskip
We apply the above lemma to the polynomial 
$P(z_0,\bz) = z_1^{N-(d+1)}$ and 
\begin{equation}\label{A2}
 \ba_n = \left( \begin{array}{c}
 -i(k_0 - p_{n0}) + \xi_{\bp_n} \\
 (\bk - 2\bp_n)/\sqrt{2m} \end{array} \right)
\end{equation}
where $\xi_{\bp} = \bp^2/(2m) - \mu$.
Since $P(1,\bz^{\bj}) = (z^{\bj}_1)^{N-(d+1)}$ and
$\ba_{j_r} \!\cdot \bz^{\bj} = 1$ for $r = 1,\dots,d+1$,
Cramer's rule yields
\begin{eqnarray}\label{A3}
 P(1,\bz^{\bj}) \prod_{n=1 \atop n \neq j_1,\dots,j_{d+1}}^N
 \det(\ba_{j_1},\dots,\ba_{j_{d+1}}) \; =
 \prod_{n=1 \atop n \neq j_1,\dots,j_{d+1}}^N z_1^{\bj} \;
 \det(\ba_{j_1},\dots,\ba_{j_{d+1}}) \; = \nonumber \\
 \prod_{n=1 \atop n \neq j_1,\dots,j_{d+1}}^N \!\!
 \det \left( \begin{array}{ccc}
  1 & \!\dots\! & 1 \\
  \frac{\bk-2\bp_{j_1}}{\sqrt{2m}} & \!\dots\! &
  \frac{\bk-2\bp_{j_{d+1}}}{\sqrt{2m}}
  \end{array} \right) = \nonumber \\
 \prod_{n=1 \atop n \neq j_1,\dots,j_{d+1}}^N
 \left( -\sqrt{2/m} \right)^d 
 \det(\bp_{j_2}\!-\!\bp_{j_1},\dots,\bp_{j_{d+1}}\!-\!\bp_{j_1})
\end{eqnarray}
In the last step we have subtracted the first column of the
determinant from all the others and then applied Laplace's
theorem.
We now evaluate the denominator in (\ref{A1}),
\begin{equation}\label{A4}
 D = \det \left( \begin{array}{cccc}
 1 & 1 & \dots & 1 \\
 -i(k_0 \!-\! p_{n0}) + \xi_{\bp_n} &
 -i(k_0 \!-\! p_{j_{1}0}) + \xi_{\bp_{j_1}} & \dots &
 -i(k_0 \!-\! p_{j_{d+1}0}) + \xi_{\bp_{j_{d+1}}} \\
 (\bk-\bp_n)/\sqrt{2m} & 
 (\bk-\bp_{j_1})/\sqrt{2m} & \dots &
 (\bk-\bp_{j_{d+1}})/\sqrt{2m}
 \end{array} \right)
\end{equation}
Subtracting the first column from all the others and applying
Laplace's theorem yields
\begin{equation}\label{A5}
 D = \det \left( \begin{array}{ccc}
 f_{j_{1}n}(\b0) & \dots & f_{j_{d+1}n}(\b0) \\
 \sqrt{2 \over m} \, (\bp_n - \bp_{j_1}) & \dots &
 \sqrt{2 \over m} \, (\bp_n - \bp_{j_{d+1}})
 \end{array} \right)
\end{equation}
Adding $\bd^{\bj} \cdot (\bp_n - \bp_{j_r})/m$
to the $r$-th matrix element in the first row (adding thus
multiples of the other rows to the first one) one obtains
\begin{eqnarray}\label{A6}
 D &=& \det \left( \begin{array}{ccc}
 f_{j_{1}n}(\bd^{\bj}) & \dots & f_{j_{d+1}n}(\bd^{\bj}) \\
 \sqrt{2 \over m} \, (\bp_n - \bp_{j_1}) & \dots &
 \sqrt{2 \over m} \, (\bp_n - \bp_{j_{d+1}})
 \end{array} \right) 
 \nonumber \\ &=&
 \left( \sqrt{2/m} \right)^d \, f_n^{\bj} \;
 \det \left( \begin{array}{ccc}
 1 & \dots & 1 \\
 \bp_n \!-\! \bp_{j_1} & \dots & \bp_n \!-\! \bp_{j_{d+1}}
 \end{array} \right)
\end{eqnarray}
Subtracting the first column from all others and applying Laplace's
theorem once again one obtains
\begin{equation}\label{A7}
 D = \left( -\sqrt{2/m} \right)^d \, f_n^{\bj} \,
 \det\big(\bp_{j_2} \!-\! \bp_{j_1},\dots,
           \bp_{j_{d+1}} \!-\! \bp_{j_1} \big)
\end{equation}
Eq.\ (\ref{A3}) and Eq.\ (\ref{A7}) yield
\begin{equation}\label{A8}
 P(1,\bz^{\bj})
 \prod_{n=1 \atop n \neq j_1,\dots,j_{d+1}}^N
 \frac{\det(\ba_{j_1},\dots,\ba_{j_{d+1}})}
      {\det \left( \begin{array}{cccc}
       1 & 1 & \dots & 1 \\
       \ba_n & \ba_{j_1} & \dots & \ba_{j_{d+1}}
       \end{array} \right) } =
 \prod_{n=1 \atop n \neq j_1,\dots,j_{d+1}}^N
 \frac{1}{f_n^{\bj}}
\end{equation}
We now set $z_0 = 0$ and $\bz = (1,\bk/\sqrt{2m})$, such that
\begin{equation}\label{A9}
 z_0 - \ba_n \!\cdot \bz = 
 i(k_0 - p_{n0}) - \xi_{\bk-\bp_n} = G_0^{-1}(k-p_n)
\end{equation}
With this choice of variables the above lemma thus yields the 
algebraic identity
\begin{equation}\label{A10}
 1 = 
 \sum_{j_1,\dots,j_{d+1} \atop 1 \leq j_1 < \dots < j_{d+1} \leq N} 
 \prod_{n=1 \atop n \neq j_1,\dots,j_{d+1}}^N
 \frac{1}{f_n^{\bj}} \, G_0^{-1}(k-p_n)
\end{equation}
Multiplying this equation with $\prod_{j=1}^N G_0(k-p_j)$ and
integrating over $k$ one finally obtains the reduction formula 
Eq.\ (\ref{eq10}).

\vfill\eject


\vspace{1cm}

\end{document}